\documentclass[preprint,12pt]{elsarticle}

\usepackage{graphicx}
\usepackage{amssymb}
\usepackage{amsmath}
\usepackage{algorithm}
\usepackage[noend]{algpseudocode}
\usepackage{caption}
\usepackage{xfrac}
\usepackage{xcolor}

\journal{JOCS}

\begin{document}

\begin{frontmatter}


\title{A Coupled lattice Boltzmann-Multiparticle collision method for multi-resolution hydrodynamics}



\author{Andrea Montessori*}

\address{Istituto per le Applicazioni del Calcolo CNR, via dei Taurini 19, Rome, Italy}

\cortext[cor1]{a.montessori@iac.cnr.it}

\author{Adriano Tiribocchi}

\address{Center for Life Nano Science@La Sapienza, Istituto Italiano di Tecnologia, 00161 Roma, Italy}

\author{Marco Lauricella}

\address{Istituto per le Applicazioni del Calcolo CNR, via dei Taurini 19, Rome, Italy}

\author{Sauro Succi}

\address{Center for Life Nano Science@La Sapienza, Istituto Italiano di Tecnologia, 00161 Roma, Italy}
\address{Institute for Applied Computational Science, John A. Paulson School of Engineering and Applied Sciences, Harvard University, Cambridge, USA}

\begin{abstract}
In this work we discuss the coupling of two mesoscopic approaches for fluid dynamics, namely the lattice Boltzmann method (LB) and the multiparticle collision dynamics (MPCD) \cite{kapral2008multiparticle} to design a new class of flexible and efficient multiscale schemes based on a dual representation of the fluid observables.

At variance with other commonly used multigrid methods, mostly oriented to high Reynolds and turbulent flows, the present approach
is  designed to capture the physics at the smallest scales whenever the lattice Boltzmann alone falls short of providing the correct physical information due to a lack of resolution, as it occurs for example in thin films between interacting bubbles or droplets in microfluidic crystals.

The coupling strategies employed to pass the hydrodynamic information between the LB and the MPCD are detailed and the algorithm is tested against the steady isothermal Poiseuille flow problem for different coupling configurations.
\end{abstract}


\end{frontmatter}



\section{Introduction}

The development of efficient and accurate numerical models able to simulate fluids' behavior over a broad range of time and spatial scales still represents a grand challenge in computational physics, with a plethora of applications in science and engineering.

Nothwistanding the huge advancements and success obtained by computer simulations at the molecular and microscopic level (i.e. MD and DSMC approaches \cite{bird1978monte,garcia1999adaptive,alder1959studies,berendsen1995gromacs}), direct simulation of complex flows at time and lenght scales of experimental relevance is still out of reach even for the most powerful supercomputers to date.

To this aim, multiscale and multiphysics approaches, able to reproduce the physical phenomena occurring at different lenght and time  scales with less computational expenditure, are constantly in demand.

Many attempts have been deployed so far in order to takle this problem and a number of concurrent hybrid approaches have been proposed to investigate a vast spectrum of physics problems, including soft matter and molecular fluids \cite{alekseeva2016hydrodynamics,potestio2013hamiltonian}, fluctuating hydrodynamics \cite{thampi2011lattice} as well as both dilute and dense hydrodynamics \cite{garcia1999adaptive,werder2005hybrid,di2016lattice}.

In these approaches, small portions of the domain are analysed at a finer scale level, whereas the remaining part is treated on a coarser and computationally less demanding level.
Then, the transfer of information enabling the coupling and communication
across the different regions takes place within a coupling region.

In this work we prospect the possibility of coupling two mesoscopic approaches for fluid dynamics, namely the lattice Boltzmann method (LB) \cite{succi2018lattice,montessori2018lattice} and the multiparticle collision dynamics (MPCD) \cite{kapral2008multiparticle} to design a novel, flexible and efficient multigrid approach. 

The lattice Boltzmann method (LB) is based on a minimal version of the Bathnagar-Gross-Krook equation \cite{bhatnagar1954model}, in which the computational molecules, namely a discrete set of probability distribution functions, propagate along the links of a cartesian lattice and collide on the nodes relaxing towards a Maxwell-Boltzmann equilibrium while, in the the MPCD,  the fluid is modeled by a collection of pointlike particles which freely stream and locally collide according to mass-momentum-energy conserving rules.

At variance with others commonly used multigrid methods, mostly oriented to high Reynolds and turbulent flows, the proposed approach
is designed to capture the physics at the smallest scales, whenever the lattice Boltzmann alone falls short of providing the correct physical information due to a lack of resolution and of underlying physics (i.e. thermal fluctuations at the micro and nano scales), as occurs for example in thin films between interacting bubbles or droplets \cite{montessori2019mesoscaleJFM,montessori2019modeling} or in micro and nano-confined flows\cite{montessori2015lattice,mohammadzadeh2013parallel}.

In the following, we provide a brief summary of the two basic methods: the LB and the MPCD.

\section{Methods}

\subsection{Lattice Boltzmann method}

The lattice Boltzmann method \cite{benzi1992} (LB) is based on a minimal version of the time-honored Bathnagar-Gross-Krook equation, in which the computational molecules, namely a discrete set of probability distribution functions, propagate along the links of a cartesian lattice and collide on the nodes relaxing towards a Maxwell-Boltzmann equilibrium.
The LB equation reads as follows:

\begin{equation}
	f_i(\vec{x}+\vec{c}_i\Delta t, t + \Delta t)=f_i(\vec{x},t) + \frac{\Delta t}{\tau} [f_i^{eq}(\rho,\vec{u}) - f_i(\vec{x},t)] 
\label{LB}
\end{equation}

where $f_i(\vec{x},t)$ is the discrete distribution function representing the probability of finding a particle at position $\vec{x}$ and time $t$ moving along the $i-th$ lattice direction with a (discrete) velocity $\vec{c}_i$, $i$ spanning over the lattice directions (i=0,...,b) \cite{succi2018lattice,montessori2018lattice}. 
For the sake of simplicity, $\Delta t$ is the lattice time step, usually set to unity.

The left hand side of the eq.\ref{LB} is the streaming step, representing the free-flight of particles along the lattice directions, which hop from a site to neighboring ones.

The right hand side is the collision term which codes for the relaxation of the set of discrete distributions towards a local equilibrium $f_i^{eq}$, namely a truncated, low Mach number expansion of the Maxwell-Boltzmann distribution \cite{succi2018lattice,kruger2017lattice}.

In this work we employed a second-order isotropic nine speed two dimensional lattice ($b=8$, $D2Q9$ lattice) equipped with a set of second-order expansion of the Maxwell-Boltzmann distribution fucntion, which read as:

\begin{equation}
f_i^{eq}= w_i\rho\left[1 + \frac{\vec{c}_i \cdot \vec{u}}{c_s^2} + \frac{(\vec{c}_i \cdot \vec{u})^2}{2 c_s^4} - \frac{\vec{u} \cdot\vec{u}}{2 c_s^2}\right]
\label{mbeq}
\end{equation} 
$w_i$ being the weights of the discrete equilibrium distribution functions, \textcolor{black}{$c_s^2 = \sum_{i} w_i {c}^{2}_{i}=1/3$} the squared lattice speed of sound, $\vec{u}$ the macroscopic fluid velocity.
Moreover, the relaxation parameter $\tau$ in equation \ref{LB} is linked to the fluid kinematic viscosity via the linear relation $\nu=c_s^2(\tau - \Delta t/2)$ \cite{succi2018lattice}.

The hydrodynamic moments located in $\vec{x}$ at time $t$, up to the order supported by the lattice in use (density, linear momentum and momentum flux tensor for a second-order isotropic lattice \cite{chen2008discrete}), can be obtained by the compution of local,linear weighted sums, namely, $\rho= \sum_{i} f_i$, $\rho \vec{u}=\sum_{i} f_i\vec{c}_i $ and $\Pi=\sum_{i} f_i(\vec{c}_i \vec{c}_i-c_s^2 \text{I})$ being $\text{I}$ the identity matrix.

\subsection{Multiparticle collision dynamics with Andersen Thermostat}

An MPCD fluid is modeled by a large number of pointlike particles of mass $m$, typically of the order of $10^{3}-10^{5}$ in two dimensions \cite{gompper2009multi,ihle2001stochastic}.
The evolution of the system during a simulation time step consists of the subsequent application of a streaming and a collision step.
During the streaming step, the particles positions are updated via a forward Euler update rule:

\begin{equation}
\vec{r}_k(t + \delta t)=\vec{r}_k(t) + \vec{v}_k(t)\delta t
\end{equation} 
being $\vec{r}_k$ the vector position and $\vec{v}_k$ the velocity of the $k-th$  particle and $\delta t$ the value of the MPCD discrete time step.

The algorithm then proceeds with the collision step, following one of the several available approaches \cite{noguchi2007particle,ihle2006consistent,allahyarov2002mesoscopic}.
Among them, a possible choice for the collision is to perform stochastic rotations of the particle velocities relative to the center-of-mass momentum.
In this case, the multiparticle collision model is usually referred to as  Stochastic Rotation Dynamics (SRD) \cite{malevanets1999mesoscopic}.

In the SRD, the domain is divided into cells of side $a$ \cite{gompper2009multi}.
Multi-particle collisions are then performed within each cell, by rotating the velocity, $\vec{\upsilon}_{k}$, of each $k$-th particle with respect to the velocity of the cell center of mass, $\vec{\upsilon}_{cm}$, of all particles in the cell:

\begin{equation}
\vec{v}_k(t+\delta t)= \vec{v}_{cm}(t) + \mathcal{R}(\vec{v}_k(t) - v_{cm}(t))
\label{srdcol}
\end{equation} 
 
In the above equation, $\mathcal{R}$ is the rotation matrix, which rotates the particles of a given cell by an angle $\pm \alpha$ with uniform probability $[\sfrac{1}{2},\sfrac{-1}{2}]$.
Because mass, momentum, and energy are conserved locally the thermohydrodynamic equations of motion are captured in the continuum limit \cite{frisch1986lattice}. 
Moreover, since the energy is locally conserved and the volume is invariant under both streaming and collision, the system is described by a microcanonical distribution at thermodynamic equilibrium \cite{gompper2009multi}.

The kinematic viscosity of an SRD fluid can be derived using a kinetic approach \cite{kikuchi2003transport} and, in two dimensions, reads as follows:
\begin{equation}
\begin{split}
\nu= \frac{N k_BT\delta t}{a^2}\left[ \frac{N}{(N - 1 + e^{-N})(1 - cos(2\alpha))}\right] + \\
\frac{m (1-cos(\alpha))}{12 \delta t}(N - 1 + e^{-N})
\end{split} 
\end{equation} 
being $N$ is the average number of particles in a collision cell.

Nonetheless, a stochastic rotation of the particle velocities relative to the center-of-mass velocity is not the only possibility to perform multi-particle collisions. 
In particular, MPCD simulations can be performed directly in the canonical ensemble by employing an Andersen thermostat (MPCD-AT) \cite{gompper2009multi,noguchi2007particle,allahyarov2002mesoscopic}.

In the MPCD-AT new relative velocities are generated during each computational step in each cell.
Thus, the collision step in a MPCD-AT can be compactly written as \cite{noguchi2007particle}:

\begin{equation}
\vec{v}_k(t+\delta t)= \vec{v}_{cm}(t) + \delta \vec{v}_k^{ran}=\vec{v}_{cm}(t) + \vec{v}_k^{ran} - \frac{1}{N_c}\sum_{j \in cell} \vec{v}_j^{ran}
\label{MPCD-AT}
\end{equation}
where $\vec{v}_k^{ran}$ are random numbers drawn from a Gaussian distribution with variance $\sqrt{k_BT/m}$ and $N_c$ is the actual number of particles in the collision cell, the sum running over all
particles in a given cell.

The MPCD-AT is, at the same time, a collision procedure and a thermostat, thus
no additional velocity rescaling is required in non-equilibrium simulations with viscous heating and, as mentioned above simulations are performed directly in the canonical ensemble \cite{gompper2009multi}.

In the MPCD with the Andersen Thermostat the kinematic viscosity takes the following form:

\begin{equation}
\begin{split}
\nu= k_BT\Delta t \left[ \frac{N}{N-1 + e^{-N}} - \frac{1}{2} \right] + \frac{a^2}{12 \Delta t} \left[ \frac{N-1 +e^{-N}}{N}  \right]
\end{split} 
\end{equation}

In this work we opted for the MPCD with the Andersen thermostat since, altough computationally more expensive than SRD, the relaxation
times in the MPCD-AT generally decrease by increasing the number of particles per cell, while they increase in the SRD. 
Longer relaxation times imply larger
number of time steps  required for transport coefficients to reach their asymptotic
values \cite{kapral2008multiparticle}. 
As a consequence the number of particles per cell commonly employed in a SRD simulation is usually confined to $5-20$, restriction which does not apply to the MPCD-AT, where the relaxation times scale as $(\ln N)^{-1}$  \cite{gompper2009multi}.

\subsection{Coupling Procedure}

In this subsection we detail the coupling strategy employed to exchange the hydrodyanmic information between the lattice Boltzmann and the multiparticle collision dynamics.

\begin{figure}
\begin{center}
\includegraphics[width=12cm]{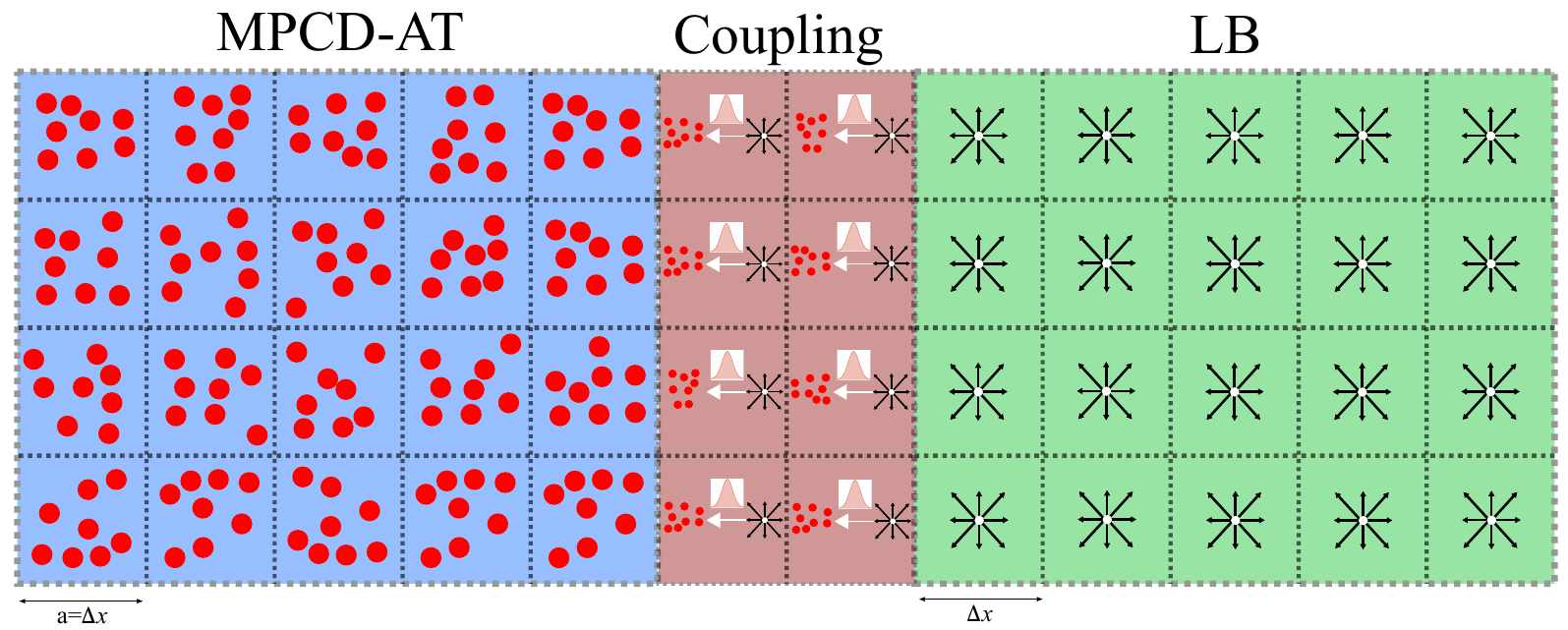}
\end{center}
\caption{Representation of the hybrid domain. On the right region, the pure LB region (green area), and on the left, the high resolution (MPCD, light blue area) region where the streaming collision process of the multiparticle collision dynamics is performed; 
in the middle, the buffer region (red area) is drawn, where the LB $\to$ MPCD step is performed via the generation of particles velocity from the LB values of the (local) linear momenta.}\label{fig:1}
\end{figure}

First, we start from the simplest coupling case, namely when $\Delta x_{LB}=\Delta x_{MPCD}=a$.
In Figure \ref{fig:1}, a sketch of the hybrid simulation domain is reported.
The right region is the pure LB region (green area), while the left one represents the high resolution region (MPCD, light blue area) where the streaming-collision processes of the multiparticle collision dynamics are performed; 
the area in the middle is the buffer region (red), where the LB-MPCD coupling  is performed via the generation of particles velocity from the LB values of the (local) linear momenta.

It is important to note that, in this work, we employ a one way coupling between the two approaches \cite{hash1996decoupled} (i.e., only LB to MPCD) since the LB runs on the entire domain, the information exchange between the two models is implemented in the coupling region while the MPCD run only in selected regions of the domain.

To be more specific, in all the simulations performed, the coupling region covers over only two lattice nodes, which has proved to be sufficient to obtain smooth transitions of the solutions between the grids.  

The coupling algorithm proceeds as follows:\\

1) \textbf{LB step}. The LB model is run across the entire domain over a computational step ( full streming and collision process ).\\

2) \textbf{Information exchange step}. In the coupling region the particles velocities are re-generated by drawing them  from a normal distribution, generated via the Box-Muller algorithm \cite{box1958note}. 
It is worth noting that, periodic boundary conditions are applied to each particle  stepping into the LB region. 
\textcolor{black}{It is worth noting that, periodic boundary conditions are applied to each particle which are in the process of  exiting from the domain. In this way we avoid to inject new particles within the MPCD domain (the mass into the MPCD domain is exactly conserved), i.e. we perform simulations in an NVT (canonical) ensamble.}\\

3) \textbf{MPCD step}. A full streaming and collision step of the MPCD-AT is performed within the MPCD region. 

\begin{figure}
\begin{center}
\includegraphics[width=12cm]{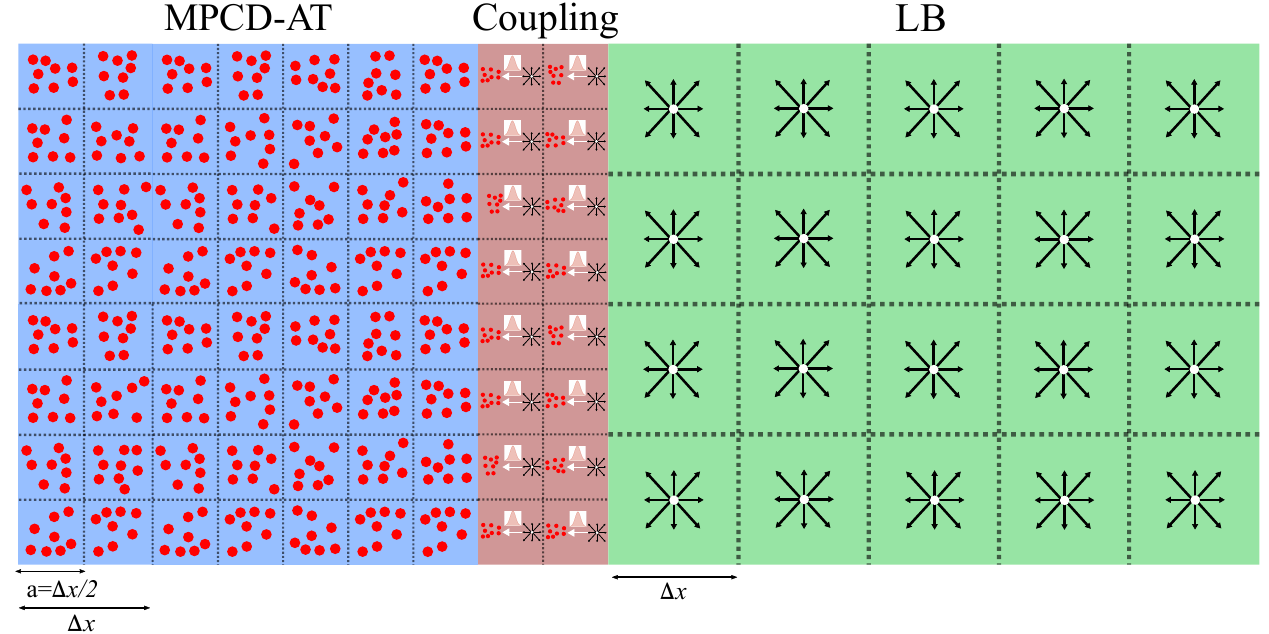}
\end{center}
\caption{Representation of the hybrid multigrid (2 level) domain.}\label{fig:2}
\end{figure}

For the sake of clarity, A pseudo-code is reported in Algorithm \ref{MPCD1}

\begin{algorithm}
\caption{Cooupling procedure : pseudo-code}\label{MPCD1}
\begin{algorithmic}[1]
\Procedure{LB-MPCD coupling}{}
\State \textbf{LB step:}\\
\State \textbf{call} LB collision
\State \textbf{call} LB streaming\\
\State \textbf{Information exchange step:}{}\\
\; \; \;\textbf{for} particles $\in$ coupling region
	draw $\vec{v}(\vec{x},t)$ from $\frac{1}{\sqrt{2 \pi k_BT}}e^{\frac{-(\vec{v} - u)^2}{2k_BT}}$\\
\State \textbf{MPCD step}{}
\State \textbf{call} MPCD streming + boundary conditions
\State \textbf{call} MPCD collision
\EndProcedure
\end{algorithmic}
\end{algorithm}

We then extended the hybrid MPCD-LB approach to run on multigrid (1 level of grid refinement) domains.  

In particular, in the present implementation, the LB code runs on the coarse grid, which extends over the whole fluid domain, while the MPCD runs on a grid with half the spacing of the LB  one(i.e. $a=1/2$, see fig. \ref{fig:2} for a visual sketch of the multigrid hybrid domain).

The implementation follows the same procedure outlined in Algorithm \ref{MPCD1} with an additional intermediate interpolation step between the LB and the MPCD steps, where the linear momentum is interpolated from the coarse to the fine grid (see Algorithm \ref{MPCD2}).

\begin{algorithm}
\caption{Cooupling procedure : pseudo-code}\label{MPCD2}
\begin{algorithmic}[1]
\Procedure{LB-MPCD two level coupling}{}
\State \textbf{LB step:}\\
\State \textbf{call} LB collision
\State \textbf{call} LB streaming\\
\State \textbf{Bilinear interpolation step:}\\
\For {$i,j$ $\in$ coupling region} $\vec{u}_{coarse} \to \vec{u}_{fine}$
\EndFor\\
\State \textbf{Information exchange step:}\\
\; \; \;\textbf{for} particles $\in$ coupling region
	draw $\vec{v}(\vec{x},t)$ from $\frac{1}{\sqrt{2 \pi k_BT}}e^{\frac{-(\vec{v} - u)^2}{2k_BT}}$\\
\State \textbf{MPCD step}{}
\State \textbf{call} MPCD streming + boundary conditions
\State \textbf{call} MPCD collision
\EndProcedure
\end{algorithmic}
\end{algorithm}

To this purpose we employed a simple bilinear interpolation.

\textcolor{black}{Summarizing, in the present coupling approach, the information is  transferred one-way only, from the LB to the MPCD domain. Consequently, the LB simulation is unaffected by the statistical thermal noise, generated within the particle domain, which is instead built-in within the MPCD method.
 It is also worth noting that, in \cite{huang2012hydrodynamic}  the equivalence between the velocity correlation function and the equipartition of kinetic energy has been demonstated, thus confirming that the fluctuations in the MPCD fluid obeys the fluctuation dissipation theorem.
As reported in Algorithms \ref{MPCD1} and \ref{MPCD2}, after a full LB step, the information is send to the MPCD domain via the generation of the particle velocities (sampled from a normal distribution) starting from the lattice Boltzmann linear momenta. 
In other words, this step works as a boundary condition for the MPCD. 
After the sampling within the buffer region, a streaming and collision step within the MPCD domain is performed.}

As pointed out above, the coupling procedure is  performed via the  generation of  the particle velocities in the overlapping region which are sampled from a Maxwell-Boltzmann distribution.
This implies that, in the coupling region, the non-equilibrium part of the velocity distribution is set to zero. In the cases investigated in this work, this particular choice turned to be effective, not compromising the accuracy of the coupling procedure, as is shown in the following.

Nonetheless,  the equilibrium coupling may become inaccurate  in the presence of strong velocity gradients, where the non-equilibrium information of the velocity distribution cannot be ignored.

In these cases, different approaches, for example the sampling of the particle velocities from an Enskog distribution \cite{garcia1998generation,distaso2016}, must be employed.  

\section{Results}

The hybrid approach was tested against the steady isothermal Poiseuille flow problem.

The flow is driven by a constant body force which acts as a  pressure gradient along the channel. 

Along the flow direction we imposed periodic boundary conditions while, on the walls, no-slip conditions are employed.

No slip conditions have been implemented with a second-order scheme (half-way bounceback \cite{kruger2017lattice}) within the LB environment while, in the MPCD, we used the method proposed by Lamura et al. in \cite{lamura2001multi}.

Three cases have been considered:

1) Standalone LB and MPCD-AT simulations. This first set up was used to calibrate and test the MPCD and to compare the two models in terms of computational time.

2) Coupled LB-MPCD using the same spatial resolution with the coupling region positioned at the center of the channel.

3) Coupled LB-MPCD with near-wall grid refinement(1 level of grid refinement).

The simulation parameters for these three sets of simulations are reported in Table \ref{Tab:1}, for the reader's convenience.

\begin{table}
\captionof{table}{Simulations parameters: $nx \times ny$ grid points,$\Delta x_{LB}$ lattice unit, $a$ bin size, $\nu_{LB}(\tau)$ LB viscosity(LB relaxation time), $\nu_{MPCD}$ MPCD viscosity, $(k_BT)$ MPCD thermal energy,$N$ MPCD particles, $N_s$ MPCD sampling interval, $g$ gravity (Body Force) \label{Tab:1}}
\setlength\tabcolsep{2pt}
\begin{tabular}{|l|l|l|l|l|l|l|l|l|}
\hline
 & $nx \times ny$ & $\Delta x_{LB}$ & $a$ & \begin{tabular}[c]{@{}l@{}}$\nu_{LB}$\\ $(\tau)$\end{tabular} & \begin{tabular}[c]{@{}l@{}}$\nu_{MPCD}$\\ $(k_BT)$\end{tabular} & $N$ & $N_s$ & $g$ \\ \hline
1)             & $30 \times 30$ &     $1$         &  $1$   &                     $0.167(1.0)$                              &                    $0.167(\sim 0.17)$                          &$60\div600$&  $100$    &  $5\cdot 10^{-4}$       \\ \hline
2)             & $30 \times 30$ &     $1$         &  $1$   &                     $0.167(1.0)$                              &                    $0.167(\sim 0.17)$                          & $1000$    &  $100$     &$5\cdot 10^{-4}$         \\ \hline
3)             & $30 \times 30$ &     $1$         &  $1/2$ &                     $0.167(1.0)$                              &                    $0.167(\sim 0.3)$                           & $1000$    &  $100$    & $5\cdot 10^{-4}$       \\ \hline
\end{tabular}
\end{table}

\subsection{ Standalone LB and MPCD-AT }

First, we run two separate sets of simulations to test the LB and MPCD-AT.

As stated above we run the classical Poisseuille flow benchmark between two parallel plates. The fluid
domain is discretized with a $30 \times 30$ nodes grid.

 As to the boundary conditions, as stated above, in both simulations we applied no slip conditions on the upper and lower walls
and periodic boundary conditions along the flow direction. 

The kinematic viscosity ($\nu$) and the body force ($g$), acting as a pressure gradient in the direction of the flow, have been set at the same values in both sets of simulations, so as to perform a one-to-one comparison between  the two models. 

All others relevant simulation parameters are reported in Table \ref{Tab:1} for the sake of clarity.
\begin{figure}
\begin{center}
\includegraphics[width=14cm]{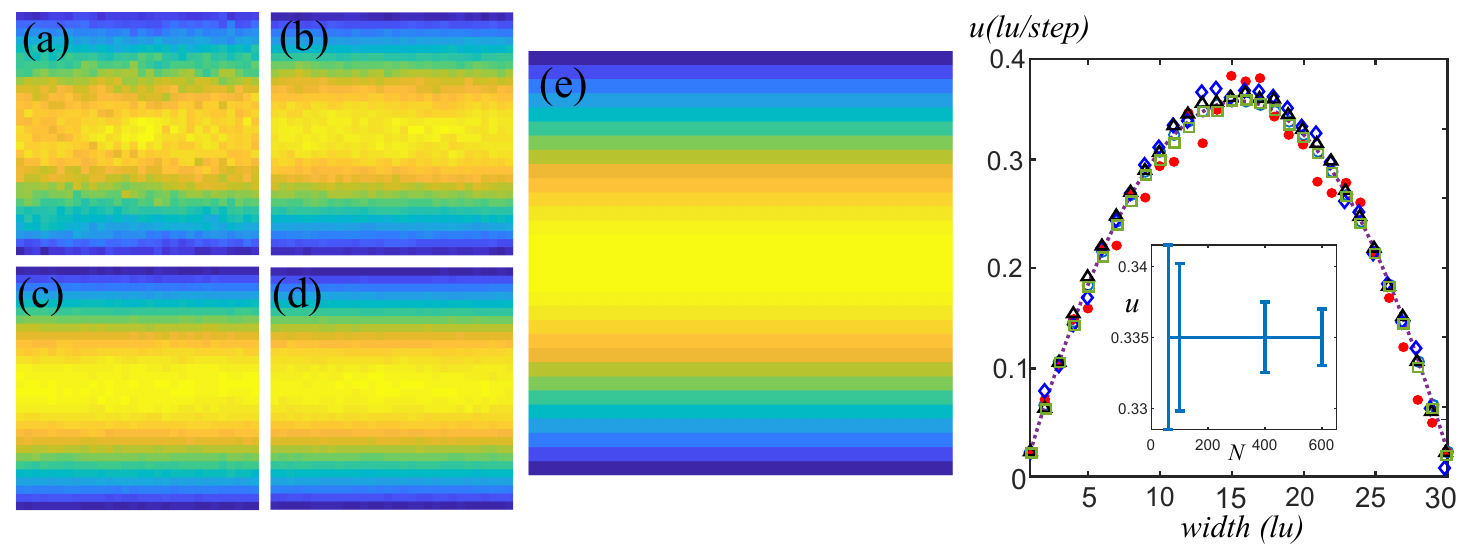}
\end{center}
\caption{(a-d) Poiseuille flow velocity field. MPCD simulations with $60$,$100$,$400$ and $600$ particles per cell. (e) Poiseuile flow velocity field for the LB model. (f) Steady state velocity profiles (dotted line(analytic solution), red circles(MPCD,$60$ particles per cell), blue diamonds(MPCD,$100$ particles per cell), black triangles(MPCD,$400$ particles per cell), green squares(MPCD,$600$ particles per cell) and hollow light blue circles(LB solution) ). Inset: standard deviation of the velocity in the centre of the channel as a function of $N$. }\label{fig:3}
\end{figure}

First we compared the steady state solutions of the MPCD and LB simulations against the analytical solution for the two-dimensional Poiseuille velocity profile.

The results, reported in figure \ref{fig:3}, confirm the accuracy of both the MPCD-AT and the LB, which are able to reproduce the Poiseuille flow with a very good degree of accuracy. 

As per the MPCD-AT, we performed four different simulations by varying the number of particle per cell between $N=60$ and $N=600$.

In each case , the velocity profiles are averaged over $N_s = 100$. 
 
In fact, a time averaging procedure is necessary to reduce the statistical noise associated with the particle nature of the MPCD. 

Furthermore, as one can see in figure \ref{fig:3}(a-d), the statistical noise decreases for increasing values of $N$. 

Indeed, as is well known, the standard deviation on the fluid velocity can be expressed (at equilibrium) as:

\begin{equation}
	\sigma_u=\sqrt{\frac{k_BT}{N}}\frac{1}{\sqrt{N_s}}
\end{equation}

where $N_s$ is the number of independent statistical samples.

From this relation the statistical error due to  the evaluation of the velocity can be estimated as follows:
\begin{equation}
	E_u= \frac{\sigma_u}{\overline{U}}
\end{equation}

which is a rough estimate of the statistical error associated to the flow field at hand. In the equation above,  $\overline{U}$ is the average value of the flow velocity.

As a check, we computed the standard deviations of the MPCD simulations reported in fig. \ref{fig:3} (a-d), which range between $0.07 \div 0.0028$ for $N=60\div 600$, thus in close agreement with the formula reported above. 
In the inset of fig.\ref{fig:3}(e), we reported the standard deviation of the velocity in the centre of the channel as a function of $N$.
\subsubsection{Computational performance}
It is now interesting to compare the computational times associated with the LB and MPCD simulations.

To start, we note that a particle update (streaming and collision) takes $ \sim 0.1 \mu s/particle/step$ which is comparable to
the time needed to update a lattice unit in a single component LB code $10 \,MLUPS$ (i.e. Mega Lattice Updates per Second), which is a typical performance of a serial implementation of a single phase, LB code \cite{kruger2017lattice}. 

It is clear that, in the case of the MPCD model, the running times scale roughly linearly with the total number of particles within the domain and, in our simulations, vary between $0.01 s/step\div 0.1 s/step$ with the particle densities ranging between $60 \div 600$ particles per bin.  

With these numbers in mind, an efficient parallelization able to exploit the computational capabilities of the latest supercomputers is needed in order to make the coupling approach amenable to large scale simulations.

It is finally worth noting that, the performance data reported above refer to a serial implementation of the code, run on a Intel Xeon Platinum 8176 CPU based on the Skylake microarchitecture. 
The code was compiled by Intel Fortran Compiler version 18.0.2 with the recent AVX-512 instruction set supported on Skylake architectures.

\subsection{ Coupled LB-MPCD-AT: center line coupling }

We then applied the coupling procedure presented above to the same test case. 

\begin{figure}
\begin{center}
\includegraphics[width=14cm]{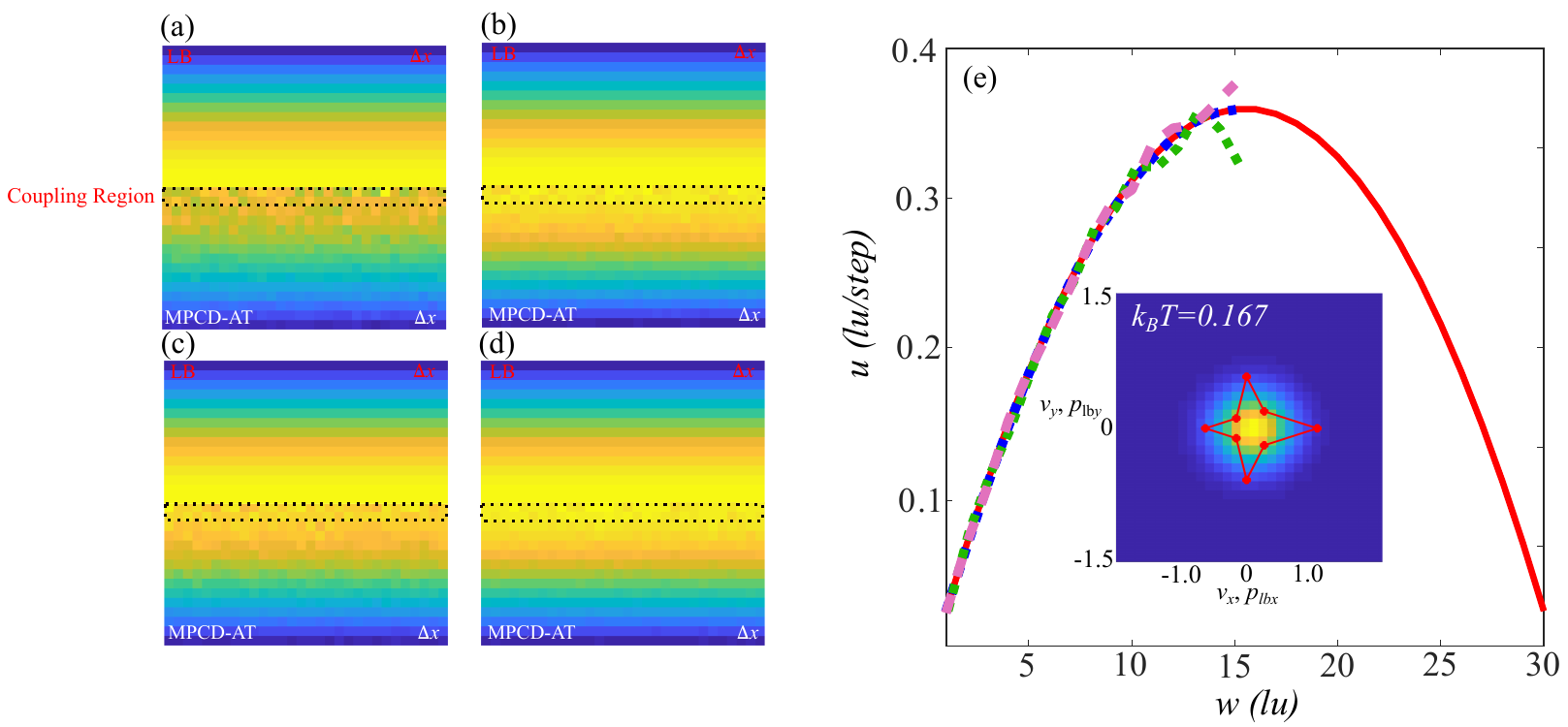}
\end{center}
\caption{(a-d) Hybrid approach, Poiseuille flow velocity field. The Coupling region is denoted by dotted region. The cell density of particles was varied between $60 \div 1000$ ((a) $60$, (b) $150$, (c) $300$, (d) $1000$). (e) Plot of the velocity profiles, LB solution (solid line) MPCD (dashed lines) for different values of $N$.  Inset: time-space averqaed velocity distribution, comparison between LB(spider plot) and MPCD ($2d$ histogram).  }\label{fig:4}
\end{figure}

The main results are shown in figure \ref{fig:4}.

As reported above, the LB run over a grid covering whole the fluid domain while the MPCD grid is restricted only to the lower half of the fluid domain.

The information is transferred at the center line of the domain, where the coupling procedure is applied. 
 
As stated above, the depth of the coupling zone, is two lattice units  and is denoted by the region included within the dashed line in figure \ref{fig:4}(a-d). 

We run four separate cases for different values of the density of MPCD particles per cell , which ranges between $60 \div 1000$ ((a) to (d)). 

The grids share the same spatial discretization and the same kinematic viscosity, $\nu=0.167\,lu^2/step$.

As one can see, the flow fields smoothly connect in the coupling region even for the smallest value of the particle density.

It is evident that, by increasing the number of particles the statistical error associated to the flow field decreases accordingly, this being  evidenced in panel (e) of the same figure which reports the LB and the MPCD velocity profiles which smoothly overlap for $N=1000$.

In the same plot,a comparison between the continuous (MPCD) and discrete
(LB) velocity distribution functions at the steady state is reported (inset of figure \ref{fig:4}(e)). 

The inset displays the time-space averaged velocity distribution.
It is interesting to note that while the MPCD velocity distribution ($2d$ field) keeps its
typical (shifted) Maxwell-Boltzmann shape the LB distributions (spider plot)
more skewed, its skewness increasing for larger values of the mean channel velocity (i.e. as the average Mach number increases).
\textcolor{black}{This departure from the Maxwell–Boltzmann behavior is due to the fact that the set of lattice distributions is not allowed to shift (i.e.,the largest probability is always associated to the rest particle), as instead occurs in the continuous case, but it occurs via a positive bias in the co-flowing distributions. }

For this reason, the larger  the non-equilibrium (represented in this case by the large values of the mean channel velocity) the larger the skewness of the discrete set of distributions.

\subsection{ Coupled LB-MPCD-AT: near-wall, two level coupling }

We then tested the capability of the hybrid approach to handle multi-level grids correctly.

To this purpose, we run two separate simulations:

1)  the LB and the MPCD share the same grid discretization. The LB runs on a $40 \times 60$ nodes grid and is coupled to the MPCD near the wall which runs on a $40 \times 14$ bins grid.

2)  the LB runs on a $20 \times 30$ nodes grid ( halved with respect to the previous case), and is coupled near the wall to the MPCD which runs on $39 \times 13$ grid with a halved discretization ($a=1/2$) with respect to the LB grid.

The first case is employed as a reference case to test the ability of the coupling procedure to correctly reproduce the Navier-Stokes solution and to handle multigrid domains.  

\begin{figure}
\begin{center}
\includegraphics[width=14cm]{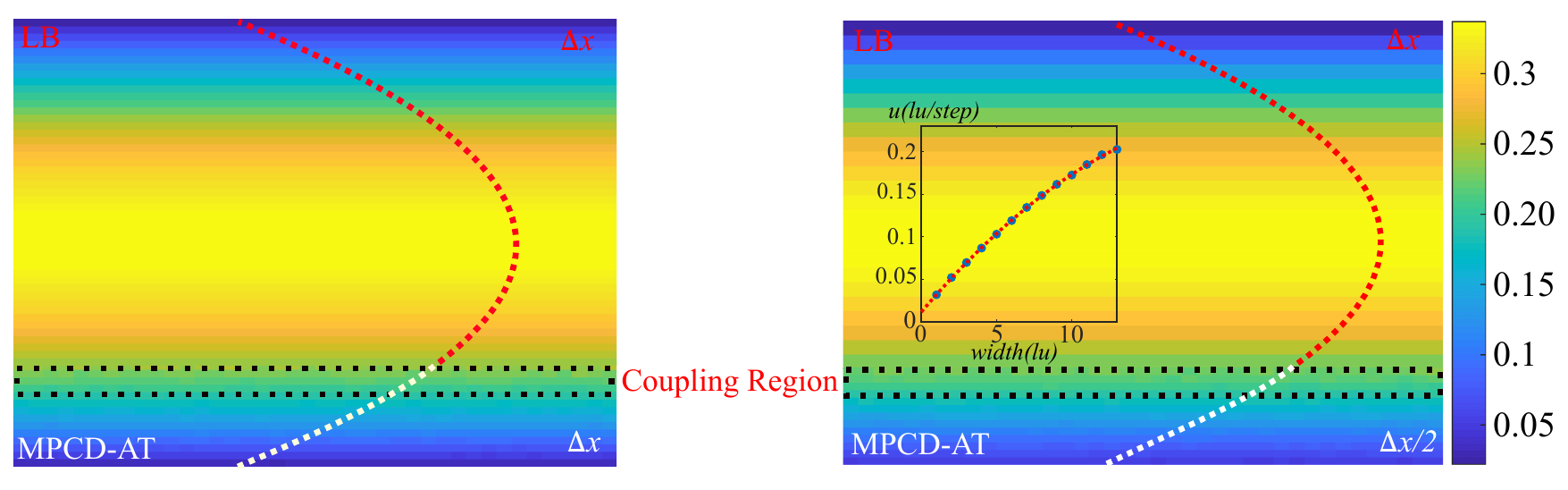}
\end{center}
\caption{(a) Coupled LB-MPCD approach with the same grid discretization (fine grid simulation). (b) Hybrid approach with grid refinement at the wall. Inset : Comparison between the analytical solution of the Poiseuille flow near the wall and the MPCD solution.}\label{fig:5}
\end{figure}

Also in this case, the overlapping zone extends to two lattice units and $N=1000 \, particles/cell$.

The main results are reported in figure \ref{fig:5}.

As one can see, the velocity profiles smoothly reconnect in the overlapping region and the number of time samples and particles per cell are sufficient to suppress the statistical oscillations inherent to the MPCD algorithm.

This is also quantitatively confirmed by the comparison between the analytical solution  near the wall and the MPCD  averaged velocity profile which perfectly overlap, as shown in the inset of fig. \ref{fig:5}, thus further corroborating the accuracy of the hybrid approach.

\section{Conclusions}

In this work we have developed a multi-level procedure to couple a lattice Boltzmann approach concurrently with a multiparticle collsion approach, the goal being to build a hybrid multigrid Navier-Stokes solver.

At variance with other commonly employed multigrid methods for high Reynolds and turbulent flows, the proposed approach
is  designed to capture the physics at the smallest scales whenever the lattice Boltzmann alone falls short of providing the correct physical information due to a lack of resolution and of underlying physics (i.e. thermal fluctuations).

The hybrid model has been applied to the classical Poiseuille flow  across two parallel plates and several cases have been investigated, namely 1) standalone LB and MPCD-AT simulations, 2) coupled LB-MPCD with the coupling region positioned at the center of the channel, 3)Coupled LB-MPCD with near-wall grid refinement(1 level of grid refinement).

Further developments of the proposed model (currently ongoing)  will aim at the simulation of complex fluid phenomena occurring in nanoconfined sysytems and also within thin films between fluid interfaces where both resolution and thermal fluctuations are required to forecast the correct physics.

\section*{Acknowledgements}

A. M., M. L., A. T. and S. S. acknowledge funding from the European Research Council under the European
196 Union’s Horizon 2020 Framework Programme (No. FP/2014-2020) ERC Grant Agreement No.739964 (COPMAT).










\end{document}